\documentclass{emulateapj}
\usepackage[T1]{fontenc}
\usepackage{textcomp}
\usepackage{amsmath}
\usepackage{mathptmx}
\usepackage{courier}
\usepackage{natbib}
\newcommand{\unitspace}{\ensuremath{\,}}
\newcommand{\usp}{\unitspace}
\newcommand{\numberspace}{\ensuremath{\;}}
\newcommand{\nsp}{\numberspace}
\newcommand{\unitstyle}[1]{\ensuremath{\mathrm{#1}}}
\newcommand{\power}[2]{\ensuremath{{#1}^{#2}}}

\newcommand{\kilo}{\unitstyle{k}}
\newcommand{\Mega}{\unitstyle{M}}

\newcommand{\cm}{\unitstyle{cm}}
\newcommand{\gram}{\unitstyle{g}}

\newcommand{\meter}{\unitstyle{m}}

\newcommand{\second}{\unitstyle{s}}

\newcommand{\Kelvin}{\unitstyle{K}}
\newcommand{\K}{\Kelvin}  

\newcommand{\grampercc}{\gram\usp\power{\cm}{-3}} 
\newcommand{\grampersquarecm}{\gram\usp\power{\cm}{-2}} 

\newcommand{\GramPerSc}{\grampersquarecm}
\newcommand{\erg}{\unitstyle{ergs}} 
\newcommand{\ergs}{\erg}

\newcommand{\amu}{\unitstyle{u}} 
\newcommand{\fermi}{\unitstyle{fm}} 
\newcommand{\eV}{\unitstyle{eV}}        
\newcommand{\keV}{\kilo\eV} 
\newcommand{\MeV}{\Mega\eV} 

\newcommand{\Msun}{\ensuremath{M_\odot}}

\newcommand{\yr}{\unitstyle{yr}}        
\newcommand{\km}{\kilo\meter}   

\newcommand{\kB}{\ensuremath{k_\mathrm{B}}} 
\newcommand{\mb}{\ensuremath{m_\mathrm{u}}} 

\newcommand{\dif}{\ensuremath{\mathrm{d}}}
\newcommand{\ee}[1]{\ensuremath{\times 10^{#1}}}

\newcommand{\satellite}[1]{\emph{#1}}

\newcommand{\beppo}{\satellite{BeppoSAX}}

\newcommand{\rxte}{\satellite{RXTE}}

\newcommand{\code}[1]{\uppercase{#1}}

\newcommand{\nuclei}[2]{\ensuremath{\mathrm{^{#1}#2}}}

\newcommand{\neutron}{n}
\newcommand{\nt}{\neutron}

\newcommand{\helium}[1][4]{\nuclei{#1}{He}}

\newcommand{\carbon}[1][12]{\nuclei{#1}{C}}

\newcommand{\titanium}[1][48]{\nuclei{#1}{Ti}}

\newcommand{\chromium}[1][52]{\nuclei{#1}{Cr}}

\newcommand{\iron}[1][56]{\nuclei{#1}{Fe}}

\newcommand{\nickel}[1][58]{\nuclei{#1}{Ni}}

\newcommand{\germanium}[1][74]{\nuclei{#1}{Ge}}

\newcommand{\krypton}[1][84]{\nuclei{#1}{Kr}}

\newcommand{\strontium}[1][88]{\nuclei{#1}{Sr}}
\newcommand{\yttrium}[1][89]{\nuclei{#1}{Y}}

\newcommand{\molybdenum}[1][98]{\nuclei{#1}{Mo}}
\newcommand{\technetium}[1][97]{\nuclei{#1}{Tc}}
\newcommand{\ruthenium}[1][102]{\nuclei{#1}{Ru}}
\newcommand{\rhodium}[1][103]{\nuclei{#1}{Rh }}
\newcommand{\palladium}[1][106]{\nuclei{#1}{Pd}}

\newcommand{\cadmium}[1][114]{\nuclei{#1}{Cd}}

\newcommand{\Ex}{\ensuremath{E_{\mathrm{exc}}}} 
\newcommand{\Ethresh}{\ensuremath{E_{\mathrm{thr}}}} 
\newcommand{\Egs}{\ensuremath{E_{\mathrm{thr,gs}\textrm{-}\mathrm{gs}}}} 
\newcommand{\mue}{\ensuremath{\mu_{e}}} 

\newcommand{\Ye}{\ensuremath{Y_e}}

\newcommand{\gamman}{\ensuremath{(\gamma,\nt)}}
\newcommand{\ngamma}{\ensuremath{(\nt,\gamma)}}
\newcommand{\source}[3]{#1~#2$#3$} 
\newcommand{\Mdot}{\ensuremath{\dot{M}}}
\newcommand{\mdot}{\ensuremath{\dot{m}}}
\newcommand{\MdotEdd}{\ensuremath{\Mdot_{\mathrm{Edd}}}}
\newcommand{\grampersecond}{\ensuremath{\gram\usp\power{\second}{-1}}}
\newcommand{\keVu}{\ensuremath{\keV\usp\amu^{-1}}}
\newcommand{\MeVu}{\ensuremath{\MeV\usp\amu^{-1}}}
\newcommand{\enu}{\ensuremath{\varepsilon_{\nu}}}
\newcommand{\ecooper}{\ensuremath{\enu^{C}}}
\newcommand{\Lnuc}{\ensuremath{L_{\mathrm{nuc}}}}
\newcommand{\yign}{\ensuremath{y_{\mathrm{ign}}}}
\newcommand{\yn}{\ensuremath{Y_{n}}}
\newcommand{\Sn}{\ensuremath{S_{n}}}
\newcommand{\omegape}{\ensuremath{\omega_{\mathrm{p},e}}}

\begin{document}

\title{Heating in the Accreted Neutron Star Ocean: Implications for Superburst Ignition}
\author{Sanjib Gupta\altaffilmark{1}, Edward F. Brown,
        Hendrik Schatz, Peter M\"oller\altaffilmark{1}, Karl-Ludwig Kratz\altaffilmark{2},\altaffilmark{3}}
\affil{Department of Physics and Astronomy, National
Superconducting Cyclotron Laboratory, and Joint Institute for
Nuclear Astrophysics,\\ 
Michigan State University, East
  Lansing, MI 48824; guptasanjib@lanl.gov, ebrown@pa.msu.edu, \\
  schatz@nscl.msu.edu, moller@lanl.gov, klkratz@uni-mainz.de}
\altaffiltext{1}{Theoretical Division, Los Alamos National Laboratory, NM 87545}
\altaffiltext{2}{Max-Planck-Institut f\"ur Chemie, Otto-Hahn-Institut, Joh.-J.-Becherweg 27, 
D-55128 Mainz, Germany}
\altaffiltext{3}{HGF VISTARS, D-55128 Mainz, Germany}

\shorttitle{Heating in Accreted Neutron Star Ocean}
\shortauthors{Gupta et al. }

\begin{abstract}
We perform a self-consistent calculation of the thermal
structure in the crust of a superbursting neutron star. In particular, we
follow the nucleosynthetic evolution of an accreted fluid element from
its deposition into the atmosphere down to a depth where the electron Fermi energy is 20~MeV. We include
temperature-dependent continuum electron capture rates and realistic
sources of heat loss by thermal neutrino emission from the crust and core.
We show that, in contrast to previous calculations, electron captures to
excited states and subsequent ${\gamma}$-emission significantly reduce
the local heat loss due to weak-interaction neutrinos. 
Depending on the initial composition these reactions release 
up to a factor of 10 times more heat at densities $<10^{11}\nsp\grampercc$ than obtained previously. This heating reduces the ignition depth of superbursts. In particular, it reduces the discrepancy noted by Cumming et al.\ between the temperatures needed for unstable \nuclei{12}{C} ignition on timescales consistent with  observations and the reduction in crust temperature from Cooper pair neutrino emission.

\end{abstract}


\section{Introduction}\label{s:introduction}

The ability to regularly monitor the X-ray sky with instruments such as
\rxte\ and \beppo\ has produced many exciting discoveries in the last few
years. An excellent example are superbursts \citep[for a review see][]{Kuulkers2003The-observers-v}.
Like
normal type I X-ray bursts \citep[for a review,
see][]{strohmayer03:review}, superbursts are characterized by a rapid rise
in the light curve followed by a quasi-exponential decay; when compared to type I
X-ray bursts, superbursts are roughly $\sim 10^{3}$ times more energetic,
have cooling timescales of hours, and recur on timescales of years. For
one source, \source{4U}{1636}{-53}, three superbursts have been observed
over a span of 4.7\nsp\yr\ \citep{wijnands:recurrent,kuulkers..ea:x-ray}.
Although most superbursts are observed at mass accretion rates
0.1--0.3\nsp\MdotEdd, with $\MdotEdd \approx 10^{18}\nsp\grampersecond$
being the Eddington accretion rate, recently
\citet{.cornelisse.ea:superbursts} detected superbursts from the rapidly
accreting source \source{GX}{17}{+2}.

The currently favored scenario for these superbursts is the thermally
unstable ignition of \carbon\ at densities $\sim 10^{8}\textrm{--}10^{9}\nsp\grampercc$
\citep{cumming.bildsten:carbon,strohmayer.brown:remarkable}.
\citet{cumming.macbeth:thermal} demonstrated that the superburst
lightcurve after the peak luminosity has been reached declines as a broken power law, 
with the early
behavior dependent on the burst energy and the later behavior dependent on
the ignition depth. This behavior has been observed
\citep{Cumming2005Long-Type-I-X-r}. The ability to constrain the burst
energetics and ignition depth is crucial, not only for understanding the
ignition of \carbon\ in a heavy-element bath, but also for probing the interior
of the neutron star
\citep{brown:superburst,cooper.narayan:theoretical,Cumming2005Long-Type-I-X-r}.
\citet{brown:superburst} demonstrated that the temperature at densities
$\sim 10^{9}\nsp\grampercc$ is sensitive to the thermal conductivity of
the inner crust and the neutrino emissivity of the core; this was
confirmed by \citet{cooper.narayan:theoretical} in a more extensive study.
Recently, \citet{Cumming2005Long-Type-I-X-r} showed that the neutrino
emission from singlet-state Cooper pairing of neutrons in the inner crust
limited the temperature at neutron drip to be $\lesssim 5\ee{8}\nsp\K$.
Taken at face value, this would lead to superburst ignition column depths
about an order of magnitude deeper than the $0.5-3\ee{12}\nsp\GramPerSc$ deduced
from observations, and correspondingly to recurrence times that are
about an order of magnitude longer than observed. This led
\citet{Page2005Superbursts-fro} to speculate that the superbursters might
in fact be strange stars.

The unstable ignition of \carbon\ depends critically
on the thermal structure of the outer crust and is sensitive to the heat
sources located there.
The outer crust of the neutron star is composed of nuclei and degenerate
electrons and exists where the electron chemical potential $\mue
\lesssim 30\nsp\MeV$ \citep[see][for a lucid discussion]{pethick98}.
The ashes of hydrogen, helium, and carbon burning
in the neutron star envelope are
continuously compressed by the ongoing accretion of matter and
incorporated into the crust. The increasing density
induces nuclear reactions at various depths that can release energy.
\emph{We show that most of this energy is deposited in the crust, rather
than being carried away by neutrinos.
This leads to a hotter crust
and decreases superburst ignition depths and recurrence times.}

\begin{figure}[tbh]
\centering{%
\includegraphics[width=3.1in]{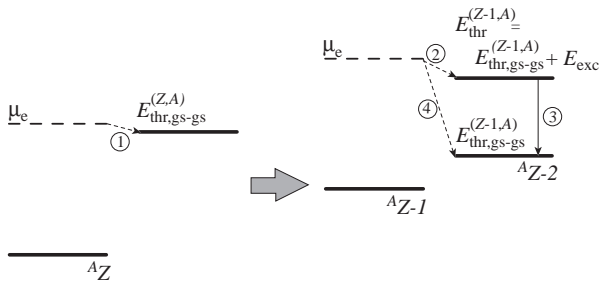}}
\caption{Electron capture for which the first
  transition (labeled ``1'') proceeds to the ground state of nucleus $^{A}Z-1$, but the second transition instead goes to an excited state ``2'' followed by a radiative de-excitation ``3''. Previous works assumed that the second transition always went to the ground state ``4''. }
\label{f:ec-schematic}
\end{figure}

Previous studies of the crust heat sources used a single
representative isotope, \iron\ \citep{sato79,haensel90a} or  \palladium[106] \citep{haensel.zdunik:nuclear}
for the product of H/He burning. \citet{haensel90a} found that
electron captures would occur in two stages (see Fig.~\ref{f:ec-schematic}). The first electron
capture (labeled ``1'' in Fig.~\ref{f:ec-schematic}) would occur 
when  $\mue\approx\Egs^{(Z,A)}$ for
capture onto an even-even nucleus: $(Z,A) + e^{-} \to
(Z-1,A)+\bar{\nu}_{e}$, where $Z$ and $A$ are even.  
Here the threshold is $\Ethresh = \Egs + \Ex$, where the threshold for the ground-state-to-ground-state transition \Egs\ is computed from atomic mass differences and therefore includes the
electron rest mass, and where \Ex\ is the energy of the lowest excited state into which the capture can proceed. Because of
the odd-even staggering of the nuclear masses, 
$\Egs^{(Z-1,A)}$ for electron capture onto the resulting
odd-odd nucleus $(Z-1,A)$ is less than $\Egs^{(Z,A)}$.
Consequently, $\mue > \Egs^{(Z-1,A)}$ and a second electron
capture $(Z-1,A)+e^{-}\to(Z-2,A)+\bar{\nu}_{e}$ immediately
follows.  Where experimental data were lacking,
which is the case for the vast majority of nuclei, the heat
deposited into the crust $Q$ was calculated assuming that the capture
was into the ground state of the daughter nucleus
(Fig.~\ref{f:ec-schematic}, label ``4''). The heat deposited into the crust is then estimated as \citep{haensel.zdunik:nuclear} $Q = (\mue - \Egs)/4$, where
the factor 1/4 arises from the integral over the reaction phase
space in the limit that $(\mue-\Ethresh)/\mue\ll 1$. This
estimate assumes that final states are very low in excitation
energy so that the contribution to the heat deposition from the
radiative de-excitation of the daughter nucleus is always
negligible compared to the heat release from the electron capture.
In this picture, 75\% of the nuclear energy release is emitted in form 
of neutrinos. 

In this paper, we calculate heating from nuclear reactions in the
outer crust (where the mass density is less than neutron drip,
$\sim 10^{11}\nsp\grampercc$) using realistic electron
capture rates obtained by microscopic calculations of the transition matrix elements between the parent ground state and daughter excited states multiplied with the phase space factor corresponding to the $Q$-value of the decay. The rates depend on the excitation energy and the ambient temperature and density. We use a full reaction network having temperature- and density- dependences. Our thermal model of the crust, described in \S~\ref{s:thermal-structure}, is coupled to the reaction network (\S~\ref{s:thermal}), so that our solution is self-consistent. Instead of
starting with a single nucleus, we follow the actual mix of nuclei initially produced
by surface burning, both from the rp-process (\S~\ref{s:rp-ash}) and a mixture in nuclear statistical equilibrium formed during a superburst (\S~\ref{s:superburst-ash}). We find that electron captures deposit considerably more energy
in the crust than in previous estimates \citep{haensel.zdunik:nuclear}. This larger heat
deposition is mainly due to the electron captures populating excited states with 
energy \Ex\ (Fig.~\ref{f:ec-schematic}, label ``2''), rather than capturing into the 
ground state as assumed previously. The populated excited states then
radiatively de-excite (Fig.~\ref{f:ec-schematic}, label ``3''), depositing the 
entire excitation energy as heat into the crust. 
We also introduce an approximate model that gives a good estimate of the heating in the 
outer regions of the crust for an arbitrary mixture in \S~\ref{s:simple-model} and describe the implications for superburst ignition in \S~\ref{s:ignition}.

\section{The Thermal Structure of the Crust}\label{s:thermal-structure}

Our thermal model is the same as used by \citet{brown:nuclear,brown:superburst} and we shall only summarize its essential properties here. Over the crust, the gravitational potential $\Phi\approx \onehalf c^{2} \ln(1-2GM/rc^{2})$ is nearly constant; moreover, we are always in hydrostatic equilibrium and can approximate the pressure as being nearly independent of temperature. Under these approximations, we then solve the equations for temperature, luminosity, and nuclear heating,
\begin{eqnarray}
e^{-\Phi/c^{2}}\frac{\dif(e^{\Phi/c^{2}} T)}{\dif r} &=& -\frac{L(1+z)}{4\pi r^{2}  K}\label{e:gradT} \\
e^{-2\Phi/c^{2}}\frac{\dif(e^{2\Phi/c^{2}}L)}{\dif r} &=& \frac{\dif L_{\mathrm{nuc}}}{\dif r}- 4\pi r^{2}(1+z)\enu, \label{e:L} \\
\frac{\dif\Lnuc}{\dif r} &=& \Mdot Q(r) \mb^{-1} \label{e:Lnuc}
\end{eqnarray}
where $z = (1-2GM/rc^{2})^{-1/2}-1$ is the gravitational redshift, $Q(r)$ is the heat deposited per accreted nucleon as computed from our reaction network (\S~\ref{s:thermal}), and $\mb=1.66\ee{-24}\nsp\gram = 1\nsp\amu$. Over the crust the potential $\Phi$ varies only slightly and our code simplifies equations (\ref{e:gradT}) and (\ref{e:L}) by holding $e^{\Phi/c^{2}}$ constant.

We model the core with an analytical density prescription $\rho = \rho_{0}[1-(r/R)^2]$ \citep{tolman:1939}, where
$\rho_{0} = 15 (8\pi)^{-1} M R^{-3}$ with $R = (2GM/c^2)
\left[1 - \left( 1+z \right)^{-2} \right]^{-1}$.  Here $r$ and $R$ are respectively the circumferential radial coordinate and stellar radius. This
prescription gives a reasonable approximation to the $\rho(r)$
that would be obtained from a modern nuclear equation of state (EOS) \citep[see][and
references therein]{lattimer.prakash:neutron}.  With this choice
for the EOS, the gravitational mass of the neutron star $M$ and redshift $z$
specify the mechanical structure of the star.  We choose the
crust-core boundary to be at a transition density
$1.6\ee{14}\usp\grampercc$, which is the density found by a
Maxwell construction using an EOS for the inner crust
\citep{negele73} and the core EOS \citep{akmal98}.  We then adopt a neutron star of mass 1.6\nsp\Msun, and set the radius at the crust-core boundary to 10.5\nsp\km. This choice makes $[1-2GM/(rc^{2})]^{-1/2} = 1.35$ at the boundary.  Our neutron star then has a radius 10.8\nsp\km\ and a surface gravity $2.43\ee{14}\nsp\cm\usp\second^{-2}$. 

For our numerical calculations we use an accretion rate, measured in the rest frame at the surface, of $\Mdot = 3.0\ee{17}\nsp\grampersecond$; the local accretion rate, per unit area, is then $2.1\ee{4}\nsp\grampersecond\usp\cm^{-2}$ and the luminosity (for spherical accretion) is 0.3 of the Eddington luminosity for a solar composition\footnote{Here $\Mdot$ is the baryon accretion rate expressed in mass units, and not the rate at which the gravitational mass of the star changes}. The core neutrino emissivity is modified Urca with an emissivity $10^{20}(T/10^{9}\nsp\K)^{8}\nsp\ergs\usp\cm^{-3}\usp\second^{-1}$. Note that because our starting density for the integration is $\rho=6.2\ee{6}\nsp\grampercc$, which is just below where the rp-process burning has concluded, we do not resolve the H/He burning shell in the envelope. Observations of the ratio of persistent fluence to burst fluence suggest that a good fraction of the accreted H/He burns stably
\citep{.cornelisse.ea:new}, so we set the temperature at this boundary to be $4.2\ee{8}\nsp\K$, which is 
computed by integrating steady H burning \citep{schatz99} at a local accretion rate of $\mdot = 2.6\ee{4}\nsp\grampersecond\usp\cm^{-2}$.

Solving equations~(\ref{e:gradT}) and (\ref{e:L}) requires the
EOS, thermal conductivity $K$ and neutrino
emissivity \enu. We use a tabulated electron EOS \citep{timmes.swesty:accuracy} with ion electrostatic interactions
\citep{farouki93}. The pressure of the free neutrons is computed
using a zero-temperature compressible liquid-drop model
\citep{mackie77}. Heat is transported by degenerate, relativistic
electrons; our treatment of the electron scattering frequency
follows \citet{brown.bildsten.ea:variability} and \citet{brown:superburst} and
we presume that the crust has a completely disordered lattice 
\citep{jones:disorder} so that the heat transport is
controlled by electron-impurity scattering \citep{itoh93}.

Our neutrino emissivity in the crust, \enu, includes contributions from bremsstrahlung \citep{haensel96}  and Cooper pairing of neutrons in the $^{1}S_{0}$ state \citep{yakovlev99:_neutr_cooper} in the inner crust.  Our treatment of neutrino-antineutrino bremsstrahlung uses a Coulomb logarithm appropriate for a liquid and does not include band-structure effects \citep{pethick94,kaminker99:_neutr}, which is consistent with our assumption that the lattice is completely disordered. In the outer crust we have neglected the plasmon neutrino emissivity \citep{itoh96:_neutr}, which is actually the dominant mode of neutrino emission there. At the temperatures of interest, however, the neutrino emission from the outer crust is negligible. 

The dominant coolant in the crust is the neutrino emission from paired neutrons in the inner crust. For the critical temperature $T_{c}$ of the paired neutrons, we choose a Gaussian in neutron Fermi momentum $k_{\mathrm{F}}$ with a maximum $\max(\kB T_{c})=0.8\nsp\MeV$ at $k_\mathrm{F} = 0.8\nsp\fermi^{-1}$,
and a width $0.28\nsp\fermi^{-1}$. These parameters are chosen to reproduce closely the
$T_{c}(k_\mathrm{F})$ of \citet{Ainsworth1989superfluid-tc}. As shown by \citet{Cumming2005Long-Type-I-X-r}, the strong temperature sensitivity of the neutrino emissivity $\ecooper$ makes the crust temperature rather insensitive to the precise dependence of the critical temperature on density. 

Recently, \citet{Leinson2006Vector-Current-} argued that vertex corrections in the expansion of the singlet-state interaction would reduce \ecooper\ by a large factor $\sim (v_{\mathrm{F}}/c)^{4}\ll 1$, where $v_{\mathrm{F}}$ is the neutron Fermi velocity. \citet{SedrakianVertex-renormal} argued that although the vertex corrections would change the temperature dependence of \ecooper, the reduction would only be of order unity. We use for \ecooper\ the rate from \citet{yakovlev99:_neutr_cooper}, but we will include, in \S~\ref{s:ignition}, a calculation with this emission suppressed to demonstrate how the ignition depth remains sensitive to \ecooper.

\begin{figure}[htb]
\centering{\includegraphics[width=3.1in]{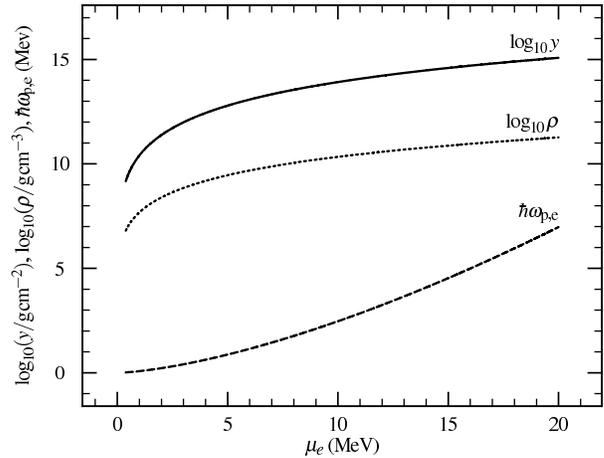}}
\caption{Column density $y = P/g$ (\emph{solid line}), mass density $\rho$ (\emph{dotted line}), and electron plasma frequency $\hbar\omegape$ (\emph{dashed line}) as a function of electron chemical potential \mue\ in the outer crust where $\mue < 20\nsp\MeV$.  The column $y$ is for a neutron star of mass $1.6\nsp\Msun$ and radius $10.8\nsp\km$.} \label{f:yRhoEf}
\end{figure}

We couple the steady-state thermal equations (\ref{e:gradT})--(\ref{e:Lnuc}) with the reaction network
calculations (\S~\ref{s:nuclear-model}) via an iterative procedure. Starting from an initial choice for the temperature profile $T(r)$, we integrate the reaction network along the path $\{T[t(r)], \rho[t(r)]\}$, where $t$ refers to the Lagrangian time since the fluid element was deposited onto the star; in our plane-parallel approximation, $t(r) = y(r)/\mdot$, with $y\approx P/g$ being the column depth. For the convenience of the reader, we show, in Figure~\ref{f:yRhoEf}, the column and mass density as a function of electron chemical potential \mue\ in the outer crust composed of rp-process ashes (\S~\ref{s:rp-ash}). The curves obey the relation $\mue\propto (\rho\Ye)^{1/3} \propto y^{1/4}$, where $\Ye$ is the electron abundance. For future reference, we also plot in Figure~\ref{f:yRhoEf} the electron plasma frequency (\emph{dashed line}), which scales with electron chemical potential as $\hbar\omegape \approx 78\nsp\keV(\mue/1\nsp\MeV)^{3/2}$.  For comparison, the thermal energies are $\kB T = 43\nsp\keV (T/5\ee{8}\nsp\K) < \hbar\omegape$ throughout much of the crust. Our reaction network does not account for this large $\hbar\omegape$ in computing \gamman\ rates.  We therefore stop the integration at  $\mue = 20\nsp\MeV$ ($\rho \gtrsim 2\ee{11}\nsp\grampercc$) where the neutron abundance begins to rise steeply. The neutrons are not yet in $\beta$-equilibrium at this point, and the density is still less than neutron drip.

The output of the reaction network gives us the local heat $Q(r)$ deposited in the shell $\{r,r+\Delta r\}$ per accreted nucleon.  We use $Q(r)$ and the abundance vector $\vec{Y}(r)$ to solve equations (\ref{e:gradT})--(\ref{e:Lnuc}) and construct a new thermal profile. For the inner crust we use the composition and heating from \citet{haensel.zdunik:nuclear}\footnote{The focus of this paper is how the thermal profile, and in particular the ignition depth of \carbon, changes when the additional nuclear physics is added to the outer crust. An exploration of the reactions in the inner crust is left for future work.}.
We then use the resulting thermal profile to update the network calculation, and we repeat this process until the thermal profile has converged.  Typically only a few iterations are required, as the heating from electron captures is insensitive to the crust temperature.

\section{Nuclear Reactions in the Outer Crust}\label{s:thermal}

The evolution of the composition and the associated nuclear energy generation, which produces the flux in the crust (eq.~[\ref{e:L}]--[\ref{e:Lnuc}]), is calculated with a nuclear reaction network of $\approx 1500$ isotopes covering the mass range $A=1\textrm{--}106$ and extending from proton-rich rp-process ashes to the neutron dripline. In this section, we first describe how the electron capture and $\beta$-decay rates are calculated. We next describe the heating in the outer crust with two different initial compositions: rp-process and superburst ashes. Building on these results, we then develop a method to approximate the heating in the outer crust for an arbitrary composition.

\subsection{The Nuclear Model}\label{s:nuclear-model}

We calculate electron capture rates for each time step of the
integration of the reaction network using a table of electron capture transition
strengths as a function of excitation energy in the daughter nucleus
and a fast analytic phase space approximation \citep{Becerril-Reyes2006Electron-Captur}
that is valid for low temperatures. This phase space calculation takes
into account the temperature and density dependence of the electron 
capture rates and is accurate enough to handle the sharp 
reaction thresholds that make table interpolation impossible. 
Temperatures and densities are calculated self-consistently
from the energy generation and composition determined by the reaction 
network using an iterative procedure with the thermal model discussed 
in \S~\ref{s:thermal-structure}.

Our nuclear model used to calculate the transition strength functions is described in 
detail in \citet{Moller1990New-development} and here we just give a summary of its properties. 
We consider only allowed Gamow-Teller transition strength, which is computed in the quasi-particle random-phase approximation (QRPA).  To obtain the wave functions we solve the Schr\"odinger equation for a deformed folded-Yukawa single-particle potential. We then add pairing and Gamow-Teller residual interactions treated in the Lipkin-Nogami and random phase approximations. Ground-state
deformation parameters and masses are obtained from the finite-range droplet model
\citep{Moller1995Nuclear-Ground-}.   We calculate neutrino losses for each transition, rather than assuming that a fixed fraction of $(\mue- \Ethresh)$ is lost to neutrinos.
As temperatures are low and radiative de-excitation
timescales are much faster than weak interaction timescales, we assume that
the parent nuclei are in their ground states. 

For the depths considered in this calculation, electron-capture-induced neutron emission is not important.
We further assume
that $\beta^-$ decay is always completely blocked. This is a good
assumption in most cases as temperatures are low, and because most
electron captures either proceed in a double step, with the second
step proceeding with $\mue \gg \Egs$, or, if they occur
at threshold, populate excited states in the final nucleus. In both
cases, \mue\ is much larger than the $\beta^-$ decay Q-value,
which is always equal or less than the ground state to ground state
electron capture threshold. We will discuss some possible exceptions to this in \S~\ref{s:rp-ash}.

While our main goal is to explore the crust heating
within the framework of electron capture rates computed from the QRPA strength functions,
we did adjust the transition energies in one case, for which
the QRPA predictions are at odds with experimental
information, and for which this has a strong effect on the
crustal heating. Our model predicts that the electron
capture on \rhodium[104] proceeds to an excited state 
at 5.3\nsp\MeV, which would result in strong heating at that depth.
Experimental data \citep{Frevert1965K-Einfang-beim-} shows, however, that while
\rhodium[104] is $\beta^-$ unstable, it also has a small ground-state-to-ground-state electron capture branch with a positive $Q$-value. As a result there is no
nuclear excitation energy released and the heat deposition from this reaction 
is much smaller than predicted from our model.

In addition to the weak interaction rates, our network includes
neutron capture rates, calculated with the statistical Hauser-Feshbach code \code{non-smoker} \citep{rauscher00}. The corresponding
\gamman\ rates are calculated from detailed balance. For consistency we use the \code{non-smoker} rates based on masses predicted by the finite-range droplet model.
For the conditions considered here the neutrons are always non-degenerate;
denoting the degeneracy by $\eta = \mu_{n}(\kB T)^{-1}$, we require for degeneracy an abundance
\begin{equation}
\yn > 0.015 \left(\eta \frac{T}{10^{9}\nsp\K}\right)^{3/2}\left(\frac{10^{12}\nsp\grampercc}{\rho}\right).
\end{equation}
For typical values $T = 5\ee{8}\nsp\K$, $\rho=2\ee{11}\nsp\grampercc$, and taking $\eta = 0.1$
as a fiducial threshold, we have $\yn > 8.4\ee{-4}$ as the abundance threshold
beyond which neutron degeneracy becomes important. At the densities we consider, the neutron abundance is well below this threshold.  Our theoretical \ngamma\ and \gamman\ rates do not account for the suppression of photons below an energy of $\hbar\omegape$ in a dense plasma. Where the neutron separation energies $\Sn > \hbar\omegape$ this does not lead to a large correction in the rate. As we show in \S~\ref{s:rp-ash}, at $\mue \gtrsim 14\nsp\MeV$ some of the nuclei present have sufficiently low \Sn\ that \gamman\ reactions would occur if the photodisintegration rates were not suppressed. A calculation of the \gamman\ and \ngamma\ rates in this regime is outside the scope of this paper; for now, we will indicate their possible impact with the recognition that a more careful treatment of these reaction rates is warranted.

The output of the reaction network is the crust composition and heat deposition $dQ = dE_{\mathrm{nuc}} + 
dE_{e} - dE_{\nu}$ for each timestep $dt = (\rho/\mdot)\,dr$. Here $dE = \sum_i dY_i \nsp BE_i$ is the nuclear energy generation calculated from the abundance change $dY_i$ and binding energy $BE_i$ of each isotope $i$, $dE_{e} = \mue d\Ye$ is the energy change due to changes in electron abundance, and $dE_{\nu}$ is the energy released as neutrinos from electron capture. Here all heat depositions are per unit mass. Thermal neutrino losses, i.e., neutrino-pair bremsstrahlung and Cooper-pair neutrino emission, are included as cooling terms \enu\ in the thermal calculation (cf.\ \S~\ref{s:thermal-structure}). We neglect the contribution from the change in Coulomb lattice energy $d E_{\mathrm{lat}}$ arising from the change in nuclear charge during an electron capture. Consider a single ion species $(Z,A)$ with number density $n_{\mathrm{ion}}$. The lattice energy is mostly due to the Madelung term, $E_{\mathrm{lat}}\approx -0.9 (n_{\mathrm{ion}}/\rho)\kB T \Gamma$, where $\Gamma  = (Ze)^{2}/(a\kB T)$ and $a$ is the  ion sphere radius.  The ratio of the lattice energy to electron energy is therefore $\propto \alpha Z^{2/3}$, where $\alpha = e^{2}/\hbar c$; with the numerical coefficients inserted this ratio is $\approx 3\ee{-3}Z^{2/3} \ll 1$ throughout the outer crust. As a result, the shift in the location of the transitions is small. Moreover, for a two-step electron capture, the contribution to $dQ$ comes from the \emph{difference} in $dE_{\mathrm{lat}}$ between the two transitions, i.e., in the relative shift of the two transitions.  This difference contributes $\approx 76\nsp\keVu (\rho/10^{10}\nsp\grampercc)^{1/3} A^{-4/3}\lesssim 1\nsp\keVu$ for the cases we consider here.

\subsection{Heat Sources in the Outer Crust} \label{s:sources}

Using the model described above, we computed the heat release for different starting compositions. In the following discussions, we highlight the dominant transitions in each.

\subsubsection{rp-Process Ashes}\label{s:rp-ash}

\begin{figure}[htpb]
\centering{\includegraphics[width=3.1in]{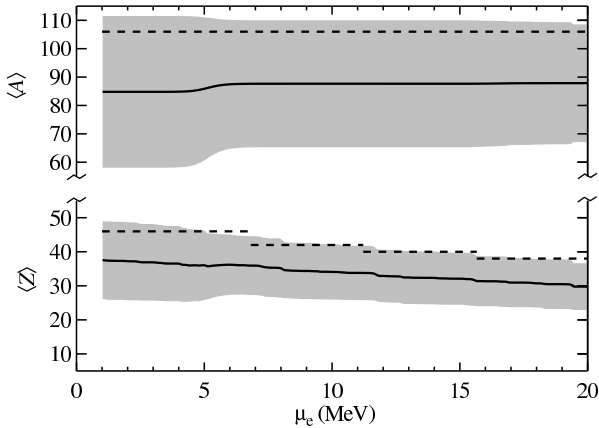}}
\caption{Composition ($\langle Z\rangle$ and $\langle A\rangle$; \emph{solid lines}) for the compression of the rp-process ashes from X-ray bursts. We show the rms deviation about $\langle Z\rangle$ and $\langle A\rangle$ (\emph{shaded bands}) for this distribution of nuclides. For comparison we also show the $Z$ and $A$ for the composition used by \citet[\emph{dashed lines}]{haensel.zdunik:nuclear}.}\label{f:compareY}
\end{figure}

We first calculate the heat sources in the outer crust using rp-process ashes from a one-zone X-ray burst model \citep{schatz.aprahamian.ea:endpoint}. The peak of the nuclide distribution is around $A = 104$, and it is useful to compare our results with those of \citet{haensel.zdunik:nuclear}, who started with a single nuclide, \palladium[106]. The initial
composition is dominated by \cadmium[104] (25\% by mass), \cadmium[105] (9.6\% by mass), and \germanium[68]
(7.1\% by mass). Other mass chains with mass fractions above 2\% are
$A=64$, 72, 76, 98, 103, and 106. There are also some small
residual amounts of \carbon\ (0.7\% by mass) and \helium\ (0.5\% by mass).
Figure~\ref{f:compareY} displays $\langle Z\rangle$ (\emph{lower solid line}) and $\langle A\rangle$ (\emph{upper solid line}) for this mixture as a function of \mue.  The rms deviations about $\langle Z\rangle$ and $\langle A\rangle$ are indicated (\emph{shaded bands}) as well. For comparison, we also plot the $Z$ and $A$ (\emph{dashed lines}) from \citet{haensel.zdunik:nuclear}.  As the fluid element chemical potential increases beyond $\mue = 4\nsp\MeV$, \carbon\ fuses into heavier
elements, with carbon being depleted by 90\% at $\mue = 5.8\nsp\MeV$ ($\rho = 4.5\ee{9}\nsp\grampercc$).  This leads to the rise in $\langle A\rangle$ shown in Figure~\ref{f:compareY}.  The abundances of \carbon\ is so small, however, that they contribute a negligible amount of heating at these shallow depths.

\begin{figure}[htpb]
\centering{\includegraphics[width=3.1in]{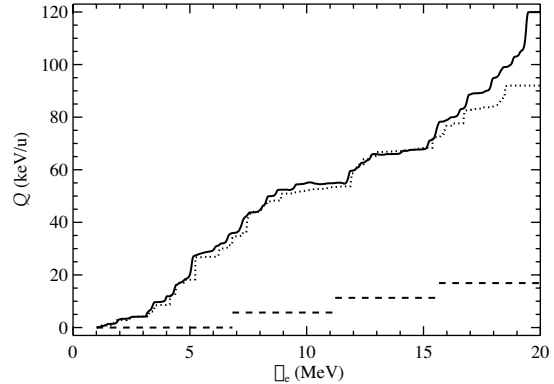}}
\caption{Integrated heat release in the crust from the compression
of the rp-process ashes from X-ray bursts (\emph{solid
line}). For comparison, we also show  (\emph{dashed line}) the integrated heat
released in the model of \citet{haensel.zdunik:nuclear}, as well as the results from our approximate
model (\emph{dotted lines}, see \S~\ref{s:simple-model}).} \label{f:compareQ}
\end{figure}

We start our calculation at $\mue = 0.38\nsp\MeV$ ($\rho = 6.2\ee{6}\nsp\grampercc$). There is an initial spike in heating as the fluid element comes into $\beta$-equilibrium. This heating is an artifact of our calculation, but, being located at the outer boundary where $T$ is fixed, does not affect the resulting thermal profile. As the fluid element is moved to greater depth, the rising \mue\ induces electron captures that reduce $\langle Z\rangle$ while preserving $\langle A\rangle$ (Fig.~\ref{f:compareY}). For $\mue \gtrsim 14\nsp\MeV$ ($\rho \gtrsim 6\ee{10}\nsp\grampercc$), $\langle A\rangle$ is no longer constant, as \gamman\ and \ngamma\ reactions alter the relative abundance of different mass chains. Figure~\ref{f:compareQ} shows the integrated deposited energy (\emph{solid line}) as a function of \mue. In this plot we set the integrated heat release to zero at $\mue = 1.0\nsp\MeV$ so that we can compare our results to those of \citet[][\emph{dashed line}]{haensel.zdunik:nuclear}.  In the range $1\nsp\MeV <\mue < 14\nsp\MeV$, electron captures deposit 66.3\nsp\keVu, which considerably exceeds the previous estimate of 16.9\nsp\keVu\ \citep{haensel.zdunik:nuclear}.  In this range of \mue, neutrinos carry away 37.2\nsp\keVu. 

This much larger heat deposition is due to our use of realistic
electron capture rates, which often lead to the population of excited
states in the daughter nucleus rather than the daughter ground
state as assumed previously. Electron capture thresholds are
therefore increased by these excitation energies. More importantly, however, the radiative de-excitation of the excited state deposits the excitation energy as heat, rather than having it carried off by neutrinos. Note that the sum of the heat deposited in the crust and the neutrino loss is 83.3\nsp\keVu, which is only slightly larger than the comparable total heat for the calculation of \citet{haensel.zdunik:nuclear}, where a single mass $A=106$ was used.

Therefore, electron captures on even-even, odd-even, or even-odd nuclei,
which tend to occur where the $\mue=\Ethresh$, can now deposit
considerable amounts of heat. These reactions were not considered
as heat sources before, as it was assumed that they proceeded from
ground state to ground state. In addition, for the electron
captures on the odd-odd nuclei---the main heat sources in previous
work---the fraction of energy lost to neutrinos is considerably
reduced.

As an example, a prominent heat source that
can be identified in Figure~\ref{f:compareQ} is the
two step electron capture transition of
\ruthenium[104] into \molybdenum[104] at $\mue=5.2\nsp\MeV$.
The electron capture on \ruthenium[104] has a threshold for
the transition into the daughter ground state of $\Egs = 5.09\nsp\MeV$
and is predicted to proceed to a low-lying state in \technetium[104] at
$\Ex = 0.15\nsp\MeV$. Therefore, this transition occurs at 
$\mue \approx \Egs + \Ex = 5.24\nsp\MeV$. The subsequent
electron capture on \technetium[104] has a lower ground state to
ground state threshold $\Egs=1.96\nsp\MeV$ and is dominated
by a transition to an excitation energy of $\Ex = 2.80\nsp\MeV$
in \molybdenum[104]. For the latter transition, about 75\% of the
transition energy $\mue - \Egs - \Ex = 0.48\nsp\MeV$ is lost to
neutrinos, but the entire excitation energy $\Ex$ for
both steps is deposited as heat, resulting in a total
heat deposition of $Q=3.07\nsp\MeV$ per transition.
The heat generated per accreted nucleon is then
$Q  Y_{104} = 8.6\nsp\keVu$, with
$Y_{104}=2.8\ee{-3}$ being the abundance in the
$A=104$ chain.  If the electron capture on \technetium[104] proceeded
through the ground state of \molybdenum[104], as assumed
in previous works, then about 75\%
of the entire transition energy $\mue - \Egs = 3.28 \nsp\MeV$ would
be lost to neutrinos and the heat deposition
would be reduced by a factor of 3.6 to
about $0.82\nsp\MeV$ per transition.
Another prominent heat source in the region $\mue \lesssim 14\nsp\MeV$ is the two-step
electron capture on \iron[68] at $\mue \approx 11.8\nsp\MeV$
(with heat deposition $Q Y_{68}=5\nsp\keVu$).

Typically, transitions in even-$A$ chains, such as
the electron capture on \ruthenium[104] discussed above,
occur as double steps
due to the odd-even staggering of the electron capture
thresholds, although
exceptions can occur if the transitions populate
high lying excited states. In contrast, thresholds
in odd-$A$ chains tend to increase steadily with electron
captures proceeding in single steps. However,
there are also some important two-step electron captures in
odd mass chains, most importantly the transition
from \rhodium[105] into \technetium[105] at $\mue =
4.4\nsp\MeV$. While the electron capture on \rhodium[105] has
a ground-state-to-ground-state threshold of 1.6\nsp\MeV, the transition
is predicted to mainly go to a 2.9\nsp\MeV\ excited state
in \ruthenium[105]. This effectively raises the threshold
to $\approx 4.5\nsp\MeV$. Therefore, even though the
electron capture on \rhodium[105] occurs at an
\mue\ near this ``effective'' threshold, the full 2.9\nsp\MeV\
excitation energy is deposited into the crust.
Following the electron capture onto \rhodium[105], \mue\ is then sufficient to
 overcome the 3.6~MeV threshold (which includes a final state excitation energy
 of 0.15~MeV) for electron capture $\ruthenium[105]
\to\technetium[105]$.
In total this sequence deposits 3.8\nsp\keVu\ into the crust ($Y_{105} = 1.3\ee{-3}$).

The mass chains $A=72$, 76, 98, and 103 also contribute 2--3\nsp\keVu\ each in the region $\mue<14\nsp\MeV$.  In total, there are about 15 transitions that generate 2\nsp\keVu\ or more and that are responsible for about 66\% of the total deposited
heat. The remainder of the heating is produced by 
a larger number of less energetic electron captures. At $\mue\gtrsim 14\nsp\MeV$, we begin to see captures of neutrons liberated by \gamman\ reactions on nuclei with low \Sn, such as \krypton[103] ($\Sn = 1.9\nsp\MeV$). As explained in \S~\ref{s:nuclear-model}, the \gamman\ rates are strongly suppressed where $\Sn < \hbar\omegape$ by the absence of photons with $\omega < \omegape$ and so the reactions we see are probably an artifact of our network. It is interesting to note, however, that in a crust composed of a mix of a number of different nuclei with different mass
numbers, the neutrons released by nuclei with low \Sn\ tend to be recaptured quickly by nuclei with
larger neutron capture $Q$-values. It is only at a larger depth, where most $A$-chains
have reached low \Sn, that an appreciable neutron abundance would begin to appear. 

\begin{figure}[htbp]
\begin{center}
\includegraphics[width=3.1in]{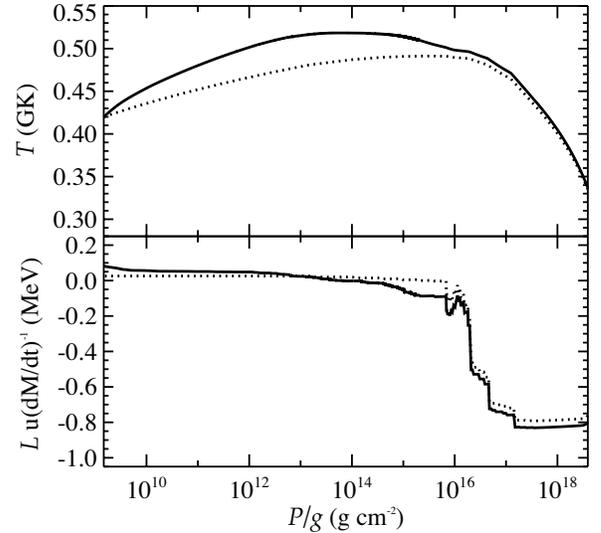}
\caption{Temperature (\emph{top}) and luminosity (\emph{bottom}) in the neutron star crust, as functions of $P/g$, for a crust composed of heavy rp-process nuclei (\emph{solid lines}). For comparison we also show the temperature and luminosity (\emph{dotted lines}) obtained using the crust composition and heating from \protect\citet{haensel.zdunik:nuclear}. We multiply the luminosity by $\mb/\Mdot$ so that the plotted quantity is in energy per accreted nucleon.}
\label{f:profile}
\end{center}
\end{figure}

Figure~\ref{f:profile} (\emph{solid lines}) demonstrates the effect on the thermal structure of the crust by the heating from all of the reactions described above. Plotted are the temperature, in units of $10^{9}\nsp\K$ (\emph{top}), and the luminosity
 $L$ (\emph{bottom}), which is multiplied by $\mb/\Mdot$ so that it is in units of MeV. For comparison, we also show (\emph{dotted lines}) the thermal profiles obtained using the composition and heating of \citet{haensel.zdunik:nuclear}. The crust temperature is clearly elevated in the outer crust.

An effect that we have not included in our calculations explicitly
and that could in principle lead to additional cooling is an Urca process that might occur for single-step, ground-state-to-ground-state electron
captures, when the subsequent electron capture is blocked until significantly
higher density. If the temperature is sufficiently high for such an electron capture to occur pre-threshold, the reverse $\beta^-$ decay
is not blocked. Until \mue\ has risen to a sufficiently
large value the electron capture occurs in tandem with a
ground-state-to-ground-state $\beta^-$ decay in the reverse direction and an Urca process will result.
While such transitions are rare (most electron captures proceed as double steps, or to excited states)
there are a few cases where they do occur, particularly in odd-mass chains far from neutron sub-shell closures.

An example is \yttrium[103], which has a ground-state threshold $\Egs=10.83\nsp\MeV$, while the electron capture daughter \strontium[103] has $\Egs=13.39\nsp\MeV$. For non-zero temperature an Urca process can occur for a small
range of $\mue$ where both \yttrium[103] electron capture and \strontium[103] $\beta^-$ decay are 
fast compared to the accretion timescale. Using the electron capture and $\beta^-$ decay rates from 
our QRPA model we estimate that only a few Urca cycles can occur for the peak temperature and accretion rate considered here. With each cycle releasing about $0.135\nsp\MeV$ per nucleus, we find for this particular 
case an additional neutrino energy loss of $0.3\nsp\keVu$ (using the abundance in the mass 103 chain 
$Y_{\rm 103}\sim7\times10^{-4}$). Even if such processes occur for about a handful of transitions, the 
total neutrino energy loss from this effect will be negligible.

The rate of energy loss from neutrinos through this mechanism is obviously very sensitive to temperature; taking a logarithmic derivative of the rate for the pre-threshold capture suggests that the neutrino loss rate is $\propto T^{7}$.  For the specific case considered here, the rate is so small that the contribution from this process is negligible for $T\lesssim 8\ee{8}\nsp\K$.
It would be interesting to explore this process in more
detail with accurate electron capture and $\beta^-$ rates for a range of temperatures
and accretion rates.

\subsubsection{Superburst ashes}\label{s:superburst-ash}

The previous calculation described the evolution of \carbon-poor ashes produced during an rp-process in an X-ray burst.  For accreting neutron stars that exhibit superbursts, which typically ignite at a density $\lesssim 10^{9}\nsp\grampercc$ \citep{Cumming2005Long-Type-I-X-r}, the burning during the superburst sets the initial crust composition rather than the X-ray burst. The composition in the region between the H/He burning shell and the carbon burning shell is also not well-described by X-ray burst ashes, as they do not contain sufficient amounts of \carbon\ to ignite superbursts. A better estimate for the initial composition in superburst systems might therefore be the ashes of \emph{stable} hydrogen and helium burning, which do
contain sufficient amounts of carbon at the accretion rates considered here
\citep{Schatz2003Nuclear-physics}. We therefore perform a second calculation,
in which we use as an initial composition the ashes of steady-state hydrogen
and helium burning at the accretion rate of $0.3\MdotEdd$
\citep{Schatz2003Nuclear-physics}. 
This ash consists mainly of $^{52}$Cr (36\% by mass), 
$^{12}$C (8\% by mass), $^{57}$Co$+^{57}$Fe (8\% by mass together), 
$^{58}$Ni (6\% by mass), and $^{60}$Ni (5\% by mass).
We then assume a composition change
at $\mue=4.3\nsp\MeV$ ($\rho = 1.7\ee{9}\nsp\grampercc$) to the superburst ashes
of \citet{schatz.bildsten.ea:photodisintegration-triggered}.
The superburst ash consists
mainly of \nickel[66] (35\% by mass), \nickel[64] (15\% by mass),
and \iron[60] (14\% by mass).

The electron captures on steady-state rp-process ashes in the region $0.45\nsp\MeV < \mue < 4.3\nsp\MeV$ result in 24\nsp\keVu\ of heating, while the superburst ashes deposit 32.5\nsp\keVu\ in the range $4.3\nsp\MeV<\mue < 14\nsp\MeV$. This latter heat deposition is slightly greater than half that deposited by the X-ray burst ashes in the same range of $\mue$.   For $\mue < 18.5\nsp\MeV$ the energy deposition is dominated by just two transitions, $\nickel[66]\to\iron[66]$ at $\mue=9.2\nsp\MeV$ and $\iron[66]\to \chromium[66]$ at $\mue=15.1\nsp\MeV$. In both these transitions
the second electron capture step populates relatively high lying excited states
at 3.6 and $2.9\nsp\MeV$ respectively, hence the relatively large heat release.
The fact that fewer transitions contribute to the heating is not unexpected as the composition is more homogeneous than in the case of the rp-process ashes.  At $\mue\gtrsim 18.5\nsp\MeV$ there is strong heating (34\nsp\keVu) from both neutron-induced reactions and the capture $\chromium[64]\to\titanium[64]$.

\subsubsection{Dependence on composition}\label{s:composition-dependence}

\begin{figure*}[htb]
\centering{\includegraphics[width=5in]{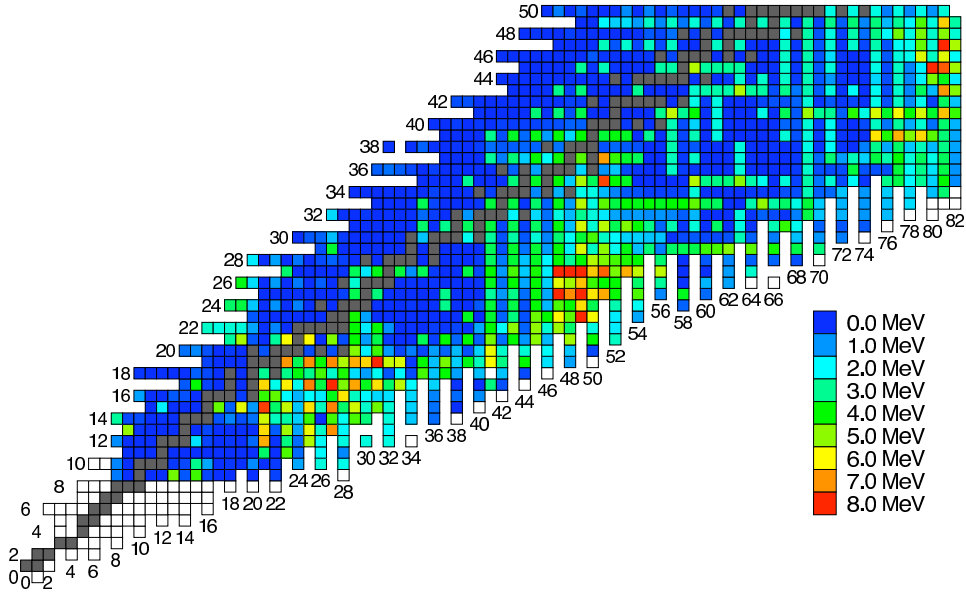}}
\caption{For each parent nucleus we show the excitation energy of
the dominant state in the daughter nucleus populated by electron
capture. Electron capture sequences passing through regions with
high excitation energies in the daughter will result in increased
heat deposition.}\label{f:exenergy}
\end{figure*}

The different heating obtained for rp-process and superburst ashes
illustrates clearly that the composition matters, when one accounts for electron captures into excited states.
The excitation energies populated by allowed Gamow-Teller
transitions depend strongly on nuclear structure, in particular on
shell and sub-shell closures. Figure~\ref{f:exenergy} shows for each
parent nucleus the main excitation energy of the electron capture
daughter nucleus. Clearly there are huge differences with particularly
high excitation energies occurring near the shell closures. If the
initial composition has a large fraction of its nuclei that will go through a region of high excitation energy, then particularly large heat deposition will occur. Away from the shell closures, nuclear
deformation tends to fragment Gamow-Teller strength and suitable
transitions are more likely to occur at lower excitation energies.

Although the \gamman\ reactions are suppressed where $\Sn < \hbar\omegape$, we note that any neutron-induced reaction sequences and their energy release will also depend strongly on composition. For example, in a mix of nuclei, \gamman\ reactions will tend to occur at shallower depths in 
$A$-chains that reach low neutron 
separation energies first (usually only one nucleus exists in a mass chain at a particular depth and the heavier, odd-mass chains lose neutrons first). 
In addition to the neutron donors, if suitable neutron-accepting nuclei are available in other mass chains at the time the neutrons are released, then a net release of heat can occur as the composition becomes more neutron bound. Depending on the composition, there will be a depth where neutron emission would occur without suitable neutron capture targets available. From there on an appreciable neutron abundance would build up.
Even if \gamman\ reactions are completely suppressed such effects might occur once the first $A$-chains have reached neutron drip deeper in the crust.
In addition, these neutron capture and \gamman\ reactions will move material to neighboring $A$-chains with different
electron capture thresholds and $\beta$-decay $Q$-values and thereby induce additional weak reactions that otherwise would not have occurred at that depth. 

\subsection{An Approximate Model}\label{s:simple-model}

To explore the composition dependence further, and to disentangle the impact of the
various differences to previous models more clearly, we present a simplified model
of the heating in the neutron star crust. In the spirit of some previous
studies \citep{haensel90a,haensel.zdunik:nuclear} we limit the model to electron captures
only (no neutron reactions), we assume that electron captures
proceed immediately when \mue\ reaches or exceeds the threshold,  and we assume
that 75\% of the electron capture transition energy is released
in form of neutrinos. In contrast to previous work, we do take
into account that electron captures populate excited states in the daughter
nuclei. We make the simplifying assumption that only the lowest lying
daughter state with significant strength is populated.

To test this model we calculate the heat release as a function of
depth for our initial rp-process ashes and compare the result (\emph{dotted line},
Figure~\ref{f:compareQ}) with the full network calculations for the rp-process burst ashes. Up to
$\mue \approx 14\nsp\MeV$ there is mostly good agreement. The slight offset at low \mue\ is because the tabulation of the approximate model does not include captures at $\mue < 2.0\nsp\MeV$. The differences at higher \mue\ are due to heating from neutron-induced reactions in the reaction network.
For captures onto nuclei for which $\mue \gg \Ethresh$ our approximate model also underestimates the fraction of heat carried off by the neutrino. Nevertheless, this approximate model provides a good estimate of the crustal heating. 

\begin{figure*}[tb]
\centering{\includegraphics[width=6.5in]{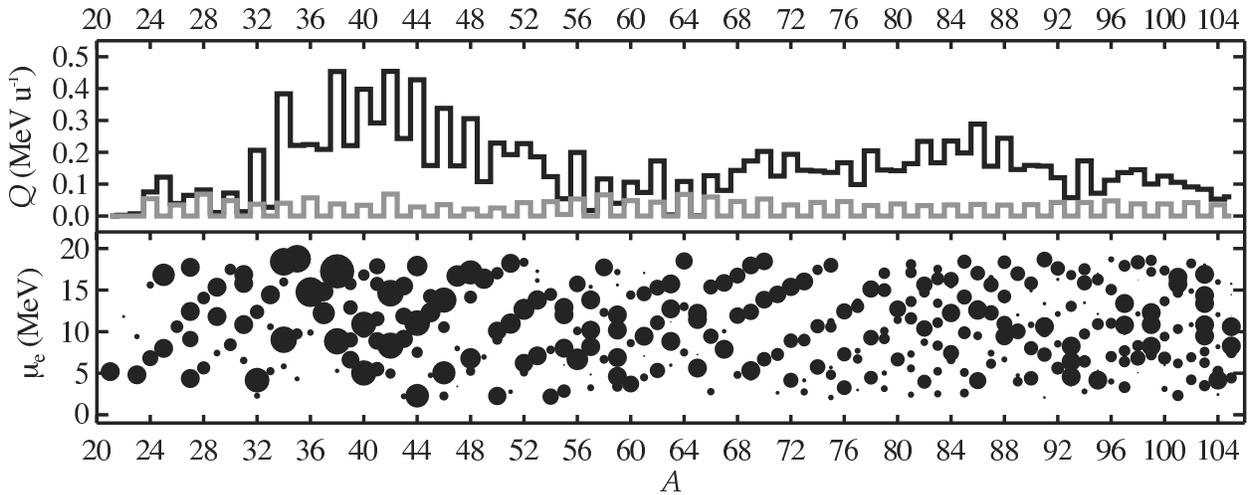}}
\caption{Electron captures in the outer crust. The mass number $A$ is along the horizontal axis; accretion pushes the fluid to greater \mue, which is upward on the bottom panel.  Each dot on the bottom plot indicates the heat deposited into the crust by a given mass chain at that \mue, with the area of each dot being proportional to the amount of heat deposited. The top panel shows the total heat deposited in the crust (\emph{black histogram}) for that mass chain, as described in \protect\ref{s:simple-model}.  In particular, the heat released by $(n,\gamma)$ reactions is neglected. For comparison, the heat that would be deposited if all transitions were ground-state-to-ground-state is also shown (\emph{top, grey histogram}).}
\label{f:captureMap}
\end{figure*}

To explore the composition dependence further, Figure~\ref{f:captureMap} maps out the heating,
 per nucleon, deposited in the crust as a function of mass number $A$
at each depth $\mue$, if the crust were composed of a single species with that mass number.
For each $A$, we start with an initial isotope $(Z,A)$ that is stable at the low-density boundary.
We then use our approximate model to
trace the heat deposition from successive electron captures with increasing $\mue$.
The area of each dot indicates the heat deposited in the crust for each transition,
and its vertical location indicates the $\mue$ at the transition.
The largest heat deposit is for $A=38$ at $\mue = 17.3\nsp\MeV$ with $Q=0.28\nsp\MeVu$.
For this calculation we only include electron captures in the region $\mue < 20\nsp\MeV$.

In Figure~\ref{f:captureMap} we also show the total deposited heat as a
function of the mass number $A$ for each mass chain (\emph{top panel, black histogram}). 
The greatest crustal heating would occur for a
composition dominated by $A=38$ or $A=42$ with heat deposition
reaching 0.45\nsp\MeVu. Heavy ashes tend to release less energy per
nucleon as excitation energies are similar but the nuclei include
more nucleons. Nevertheless, shell effects lead to large heating
for $A=68\textrm{--}91$ with particularly high values for $A=70,$ 84, 86, and
88. In systems without superbursts, rp-process ashes rich in such
nuclei would lead to a hotter crust. Superburst ashes in the mass
range $A=60\textrm{--}66$ and the rp-process ashes dominated by $A=104$ are less
favorable for heating the crust.

For comparison, we also show the heat deposition calculated with the
same method but assuming ground-state transitions only (\emph{grey histogram}). Taking into account
realistic transition energies for electron captures leads to a dramatic increase in
heat deposition that is strongly composition dependent.

\section{Superburst Ignition}\label{s:ignition}

\begin{figure}[htbp]
\centering{\includegraphics[width=3.1in]{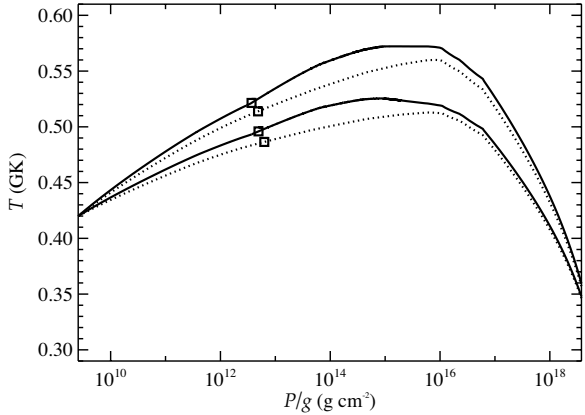}}
\caption{Thermal profiles for crusts with an outer layer of \emph{steady state} rp-process ashes, an outer crust composed of superburst ashes, with heating computed either by our network calculations (\emph{solid lines}) or according to the model of \protect\citet[][\emph{dotted lines}]{haensel90a}, and an inner crust computed from the model of \protect\citet{haensel90a}.  The upper set of curves have the Cooper-pair neutrino emissivity suppressed.  The column where \carbon\ unstably ignites is indicated for each solution (\emph{squares}).}
\label{f:ignition}
\end{figure}

As shown in Figure~\ref{f:compareQ}, the heat deposited in the outer crust is roughly 4 times
greater than that computed in an approach that neglects captures into excited states.
This greater heat release in the outer crust raises the temperature (Fig.~\ref{f:profile}) in the region where \carbon\ would unstably ignite, and hence one signature of this enhanced heating is a reduction in the ignition column depth \yign\ for superbursts. We calculate the column depth \yign\ for unstable
carbon ignition using the composition from steady-state rp-process burning and the superburst ashes as described in \S~\ref{s:superburst-ash}. These ashes do contain \carbon\ at mass fraction $X(\carbon) =0.08$. After solving for the thermal profile (Fig.~\ref{f:ignition}), as described in \S~\ref{s:thermal-structure}, we locate the ignition column \yign\ using the one-zone stability analysis described in \citet{cumming.bildsten:carbon}. Our reaction rate is taken from \citet{caughlan88:_therm}, with screening according to the prescription of \citet{ogata93:_therm}.  These results overestimate the enhancement of the reaction rate \citep{Ogata1997Enhancement-of-,Yakovlev2006Fusion-reaction}. For the case of $\carbon+\carbon$ fusion and the densities of interest, we find that our ignition depths increase by $\lesssim 20\%$ when using the screening term indicated by \citet{Yakovlev2006Fusion-reaction}.

These ignition points are marked (\emph{squares}) on Figure~\ref{f:ignition}.  The ignition column for the four models, from shallowest to deepest, are $\yign/(10^{12}\nsp\GramPerSc) = 3.7$, 4.8, 4.9, and 6.3. The corresponding recurrence times are $t_{\mathrm{rec}} = \yign / \mdot = 5.6$, 7.2, 7.4, and 9.5\nsp\yr.  The change in outer crust composition from a single mass $A = 56$ with only captures into ground state to a realistic mixture of rp-process and superburst ashes shortens the recurrence time by about 20\%.
The increase in temperature at $y \lesssim 10^{12}\nsp\GramPerSc$ is much less dramatic between the two crust models. This is due in part because both the steady-state rp-process and the superburst produce mass chains that are devoid of captures at low \mue\ (cf.\ Fig.~\ref{f:captureMap}). 

\begin{figure}[htbp]
\begin{center}
\includegraphics[width=3.1in]{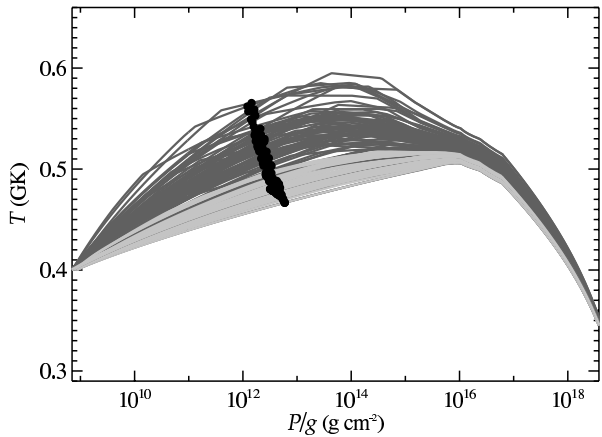}
\caption{Temperature in the neutron star crust for mass chains with $A=21\textrm{--}105$. We show two models, one with captures into excited states included (\emph{dark grey lines}) and one with only ground-state captures (\emph{light grey lines}).
In each case we set $\Mdot=3.0\ee{17}\nsp\gram\usp\second^{-1}$ and impose the same boundary conditions.  For comparison, we also show for each model where \carbon\ would unstably ignite (\emph{dots}).}
\label{f:temperature}
\end{center}
\end{figure}

We use the approximate model (\S~\ref{s:simple-model}) to investigate how the ignition depth depends on the heating in, and hence composition of, the outer crust. For each mass chain, we computed a thermal profile for a crust composed of that nuclide. As shown in Figure~\ref{f:temperature}, there is a wide variety in temperatures in the outer crust. The cases shown have the same boundary conditions as described in \S~\ref{s:thermal-structure}. As Figure~\ref{f:captureMap} shows there is a wide variation among the different $A$-chains in the amount of heat deposited and the deposition depth, which leads to a sizable difference in temperature at column $y\sim 10^{12}\nsp\GramPerSc$.

\begin{figure}[htbp]
\begin{center}
\includegraphics[width=3.1in]{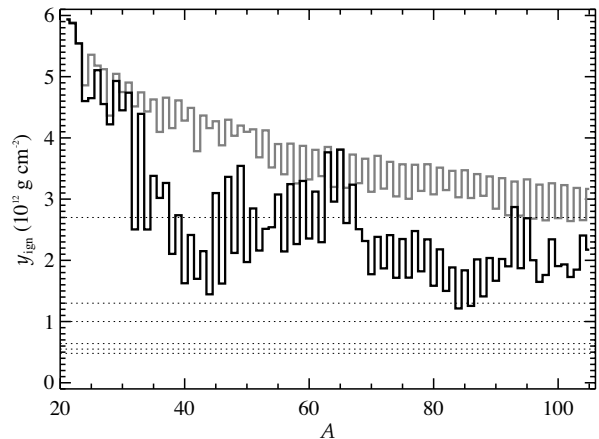}
\caption{Carbon ignition depth \yign\ for crusts composed of different mass chains $A$. The thermal profile is calculated using our approximate model.  The difference in heating between the cases for
which captures into excited states are included (\emph{black
histogram}) and neglected (\emph{grey histogram}) are clearly
evident. For comparison, we also show the fitted ignition depths according to \citet[Table 1]{Cumming2005Long-Type-I-X-r} (\emph{thin dotted lines}).} \label{f:recurrence}
\end{center}
\end{figure}

To follow how this variation in temperature affects the ignition depth, we then computed where a mixture of the the crust material and \carbon, with $X(\carbon) = 0.2$, would unstably ignite.  The ignition points are shown in Figure~\ref{f:temperature} (\emph{black dots}), and we plot $\yign$ as a function of the mass chain $A$ in Figure~\ref{f:recurrence} (\emph{black histogram}).  For
comparison, we also show the ignition depth computed when captures
into excited states are neglected (\emph{grey histogram}). 
There is a general trend that the ignition depth decreases with increasing $A$, even for the case with no
excitation energy. This is a consequence of two factors. First, the electron density at a given column (pressure) 
increases slightly with $\langle Z\rangle$ to offset the negative Coulomb contribution to the pressure. Second, the
timescale for a temperature perturbation at a column of
$10^{12}\nsp\GramPerSc$ to diffuse away increases by a factor of
$\approx 3$ as $A$ increases from 21 to 105.  Both of these
effects act to reduce the column needed to unstably ignite a
\nuclei{12}{C}-enriched plasma.  This variation with $A$ is probably exaggerated somewhat, because we did not self-consistently solve the thermal profile with the admixture of \carbon\ at $y < \yign$, which would reduce the disparity in the mean value of $Z$ and $A$ among mass chains. Nevertheless, this exercise demonstrates how the superburst ignition depth depends on the makeup of the ashes of H and He burning, and the importance of correctly treating the underlying nuclear physics of the crust.  The range of ignition columns is more consistent with the fitted recurrence times (Fig.~\ref{f:recurrence}, \emph{thin dotted lines}) of \citet[Table 1]{Cumming2005Long-Type-I-X-r} when captures into excited states are included, but is still not shallow enough to accommodate all of the inferred ignition depths.  To reduce the ignition column  further would require a reduction in the core neutrino emissivity below modified Urca, as suggested by \citet{Cumming2005Long-Type-I-X-r}.

\begin{figure}[htbp]
\begin{center}
\includegraphics[width=3.1in]{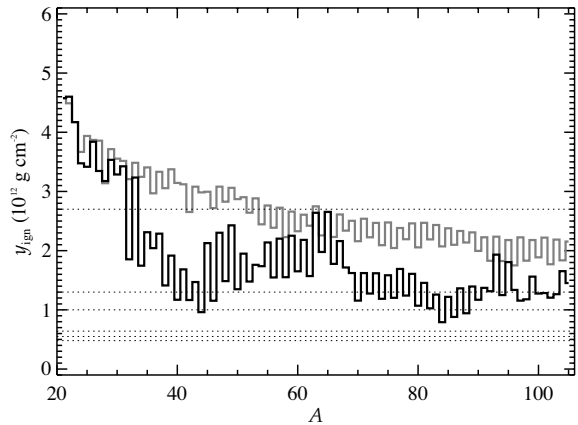}
\caption{Same as Fig.~\protect\ref{f:recurrence}, but for the Cooper-pairing neutrino emissivity suppressed.} \label{f:recurrence-Coop}
\end{center}
\end{figure}

If the Cooper-pairing neutrino emissivity were suppressed in the inner crust, then the ignition depth is decreased further (Fig.~\ref{f:recurrence-Coop}).  As before, the inclusion of electron captures into excited states reduces the discrepancy between the calculated \yign\ and those fitted from observations.

\section{Summary and Conclusions}\label{s:conclusions}

We show that allowing for electron captures into excited states drastically increases the heat deposited into the outer crust of accreting neutron stars. The fraction of the reaction energy carried away by neutrinos is sharply curtailed.  We also find that the amount of heating in the crust is very sensitive to the initial composition forged during the burning of light elements in the neutron star envelope. This sensitivity is a consequence
of the pronounced shell and sub-shell structure of neutron-rich nuclei, which can lead to drastic changes in shape, single-particle level structure, and electron capture strength distributions, even between nuclei of similar $Z, A$.

We have explored how this increased deposition affects the ignition depth, and hence the recurrence time, of superbursts. For a realistic composition---steady-state rp-process ashes overlying superburst ashes---the inclusion of the excited states decreases the ignition depth by about 20\%.  For the most optimistic case we consider, in which the neutrino emissivity from Cooper-pairing is suppressed, the ignition depth for  accretion at 0.3 Eddington is $\yign=3.7\ee{12}\nsp\GramPerSc$, closer to the value inferred from observations ($\yign = 5\ee{11}\textrm{--}2.7\ee{12}\nsp\GramPerSc$; \citealt{Cumming2005Long-Type-I-X-r}), although a significant discrepancy remains unless the core neutrino emissivity is reduced below the modified Urca rate.

To explore the variation of superburst ignition depths with composition, we have constructed an approximate model of the heating in the crust. Using this model and including the neutrino emissivity from paired neutrons, we find that at 0.3 Eddington the superburst ignition depth can be as low as $1.3\ee{12}\nsp\GramPerSc$ if the composition is dominated by nuclei with $A=40\textrm{--}44$ or 82--86. The corresponding recurrence time for ignition at this depth, with the assumed accretion rate and neutron star parameters, is 2\nsp\yr. This is roughly consistent with the observations, without invoking any additional physics. We are not aware, however, of a mechanism that would, for example, avoid the production of nuclei around $A \approx 60$ dominating the superburst ashes.

Our results also indicate that crustal heating is very sensitive to the underlying nuclear physics. Reliable estimates of the structure of very neutron-rich nuclei up to $A=106$ are needed to constrain the lowest lying electron capture strength, which determines the heating from the deexcitation of excited daughter states.  Nuclear masses are needed to determine reaction thresholds and neutron separation energies. Finally, masses, $\beta^-$decay rates, and reaction rates for the proton- and helium-induced reactions on very neutron deficient nuclei during hydrogen and helium burning in the $\alpha p$- and rp-processes are needed to accurately determine the initial composition. More work is also needed to extend our calculations into the inner crust to determine the heating there.

\acknowledgements

It is a pleasure to thank Lars Bildsten, Andrew Cumming, Dany Page, and Andrew Steiner for
stimulating and fruitful conversations, and the referee for a careful and through reading of the text, which greatly improved the manuscript. We thank
Friedel Thielemann for providing the network solver and Michelle Ouellette
for contributing to the network code. This work is supported by the
Joint Institute for Nuclear
Astrophysics under NSF-PFC grant PHY~02-16783 and by NSF grant PHY~0110253. K. L. K. acknowledges support from Virtuelles Institut f\"ur Struktur der Kerne und nukleare Astrophysik (VISTARS) under HGF grant VH-VI-061. This work was partially carried out under the auspices of the National Nuclear
Security Administration of the US Department of Energy at Los Alamos National Laboratory under Contract  DE-AC52-06NA25396.

\bibliographystyle{apj}
\bibliography{master}

\begin{thebibliography}{49}
\expandafter\ifx\csname natexlab\endcsname\relax\def\natexlab#1{#1}\fi

\bibitem[{Ainsworth {et~al.}(1989)Ainsworth, Wambach, \&
  Pines}]{Ainsworth1989superfluid-tc}
Ainsworth, T.~L., Wambach, J., \& Pines, D. 1989, Phys. Lett. B, 222, 173

\bibitem[{Akmal {et~al.}(1998)Akmal, Pandharipande, \& Ravenhall}]{akmal98}
Akmal, A., Pandharipande, V.~R., \& Ravenhall, D.~G. 1998, \prc, 58, 1804

\bibitem[{{Becerril Reyes} {et~al.}(2006){Becerril Reyes}, Gupta, Kratz,
  M\"oller, \& Schatz}]{Becerril-Reyes2006Electron-Captur}
{Becerril Reyes}, A.~D., Gupta, S., Kratz, K.~L., M\"oller, P., \& Schatz, H.
  2006, in Nuclei in the Cosmos IX, ed. J.~{Cederk{\"a}ll} {et~al.} (Trieste:
  SISSA)

\bibitem[{{Brown}(2000)}]{brown:nuclear}
{Brown}, E.~F. 2000, \apj, 531, 988

\bibitem[{Brown(2004)}]{brown:superburst}
Brown, E.~F. 2004, \apjl, 614, L57

\bibitem[{{Brown} {et~al.}(2002){Brown}, {Bildsten}, \&
  {Chang}}]{brown.bildsten.ea:variability}
{Brown}, E.~F., {Bildsten}, L., \& {Chang}, P. 2002, \apj, 574, 920

\bibitem[{Caughlan \& Fowler(1988)}]{caughlan88:_therm}
Caughlan, G.~R. \& Fowler, W.~A. 1988, At.\ Data Nucl.\ Data Tables, 40, 283

\bibitem[{Cooper \& Narayan(2005)}]{cooper.narayan:theoretical}
Cooper, R.~L. \& Narayan, R. 2005, \apj, 629, 422

\bibitem[{Cumming \& Bildsten(2001)}]{cumming.bildsten:carbon}
Cumming, A. \& Bildsten, L. 2001, \apj, 559, L127

\bibitem[{{Cumming} \& {Macbeth}(2004)}]{cumming.macbeth:thermal}
{Cumming}, A. \& {Macbeth}, J. 2004, \apjl, 603, L37

\bibitem[{Cumming {et~al.}(2006)Cumming, Macbeth, in~'t Zand, \&
  Page}]{Cumming2005Long-Type-I-X-r}
Cumming, A., Macbeth, J., in~'t Zand, J. J.~M., \& Page, D. 2006, \apj, 646,
  429

\bibitem[{Farouki \& Hamaguchi(1993)}]{farouki93}
Farouki, R. \& Hamaguchi, S. 1993, \pre, 47, 4330

\bibitem[{Frevert {et~al.}(1965)Frevert, Schoneberg, \&
  Flammersfeld}]{Frevert1965K-Einfang-beim-}
Frevert, L., Schoneberg, R., \& Flammersfeld, A. 1965, Z. Phys., 185, 217

\bibitem[{{Haensel} {et~al.}(1996){Haensel}, {Kaminker}, \&
  {Yakovlev}}]{haensel96}
{Haensel}, P., {Kaminker}, A.~D., \& {Yakovlev}, D.~G. 1996, \aap, 314, 328

\bibitem[{{Haensel} \& {Zdunik}(1990)}]{haensel90a}
{Haensel}, P. \& {Zdunik}, J.~L. 1990, \aap, 227, 431

\bibitem[{{Haensel} \& {Zdunik}(2003)}]{haensel.zdunik:nuclear}
---. 2003, \aap, 404, L33

\bibitem[{{in 't Zand} {et~al.}(2004{\natexlab{a}}){in 't Zand}, Cornelisse, \&
  Cumming}]{.cornelisse.ea:superbursts}
{in 't Zand}, J. J.~M., Cornelisse, R., \& Cumming, A. 2004{\natexlab{a}},
  \aap, 426, 257

\bibitem[{{in 't Zand} {et~al.}(2004{\natexlab{b}}){in 't Zand}, Cornelisse,
  Kuulkers, Verbunt, \& Heise}]{.cornelisse.ea:new}
{in 't Zand}, J. J.~M., Cornelisse, R., Kuulkers, E., Verbunt, F., \& Heise, J.
  2004{\natexlab{b}}, in Proc. "X-Ray Timing 2003: Rossi and Beyond", ed.
  J.~H.~S. P.~Kaaret, F. K.~Lamb (Melville, NY: American Institute of Physics),
  in press (astro-ph/0407087)

\bibitem[{Itoh {et~al.}(1996)Itoh, Hayashi, Nishikawa, \&
  Kohyama}]{itoh96:_neutr}
Itoh, N., Hayashi, H., Nishikawa, A., \& Kohyama, Y. 1996, \apjs, 102, 411

\bibitem[{Itoh \& Kohyama(1993)}]{itoh93}
Itoh, N. \& Kohyama, Y. 1993, \apj, 404, 268

\bibitem[{Jones(2004)}]{jones:disorder}
Jones, P.~B. 2004, \prl, 93, 221101

\bibitem[{Kaminker {et~al.}(1999)Kaminker, Pethick, Potekhin, Thorsson, \&
  Yakovlev}]{kaminker99:_neutr}
Kaminker, A.~D., Pethick, C.~J., Potekhin, A.~Y., Thorsson, V., \& Yakovlev,
  D.~G. 1999, \aap, 343, 1009

\bibitem[{Kuulkers(2004)}]{Kuulkers2003The-observers-v}
Kuulkers, E. 2004, Nucl. Phys. B Proc. Suppl., 132, 466

\bibitem[{Kuulkers {et~al.}(2004)Kuulkers, {in 't Zand}, Homan, {van Straaten},
  Altamirano, \& {van der Klis}}]{kuulkers..ea:x-ray}
Kuulkers, E., {in 't Zand}, J., Homan, J., {van Straaten}, S., Altamirano, D.,
  \& {van der Klis}, M. 2004, in X-ray Timing 2003: Rossi and Beyond, ed.
  P.~Kaaret, F.~K. Lamb, \& J.~H. Swank (Melville, NY: AIP Press)

\bibitem[{{Lattimer} \& {Prakash}(2001)}]{lattimer.prakash:neutron}
{Lattimer}, J.~M. \& {Prakash}, M. 2001, \apj, 550, 426

\bibitem[{Leinson \& Perez(2006)}]{Leinson2006Vector-Current-}
Leinson, L.~B. \& Perez, A. 2006, Phys. Lett. B, 638, 114

\bibitem[{Mackie \& Baym(1977)}]{mackie77}
Mackie, F.~D. \& Baym, G. 1977, \nphysa, 285, 332

\bibitem[{{M{\"o}ller} {et~al.}(1995){M{\"o}ller}, {Nix}, {Myers}, \&
  {Swiatecki}}]{Moller1995Nuclear-Ground-}
{M{\"o}ller}, P., {Nix}, J.~R., {Myers}, W.~D., \& {Swiatecki}, W.~J. 1995,
  At.\ Data Nucl.\ Data Tables, 59, 185

\bibitem[{M{\"o}ller \& Randrup(1990)}]{Moller1990New-development}
M{\"o}ller, P. \& Randrup, J. 1990, \nphysa, 514, 1

\bibitem[{Negele \& Vautherin(1973)}]{negele73}
Negele, J.~W. \& Vautherin, D. 1973, \nphysa, 207, 298

\bibitem[{{Ogata}(1997)}]{Ogata1997Enhancement-of-}
{Ogata}, S. 1997, \apj, 481, 883

\bibitem[{Ogata {et~al.}(1993)Ogata, Ichimaru, \& {Van Horn}}]{ogata93:_therm}
Ogata, S., Ichimaru, S., \& {Van Horn}, H.~M. 1993, \apj, 417, 265

\bibitem[{Page \& Cumming(2005)}]{Page2005Superbursts-fro}
Page, D. \& Cumming, A. 2005, \apjl, 635, L157

\bibitem[{Pethick \& Ravenhall(1998)}]{pethick98}
Pethick, C.~J. \& Ravenhall, D.~G. 1998, in Trends in nuclear physics, 100
  years later, ed. H.~e.~a. Nifenecker, Les Houches Sesion LXVI, 1996 (New
  York: Elsevier), 717

\bibitem[{{Pethick} \& {Thorsson}(1994)}]{pethick94}
{Pethick}, C.~J. \& {Thorsson}, V. 1994, \prl, 72, 1964

\bibitem[{{Rauscher} \& {Thielemann}(2000)}]{rauscher00}
{Rauscher}, T. \& {Thielemann}, F. 2000, At.\ Data Nucl.\ Data Tables, 75, 1

\bibitem[{{Sato}(1979)}]{sato79}
{Sato}, K. 1979, Prog.\ Theor.\ Physics, 62, 957

\bibitem[{{Schatz} {et~al.}(2001){Schatz}, {Aprahamian}, {Barnard}, {Bildsten},
  {Cumming}, {Ouellette}, {Rauscher}, {Thielemann}, \&
  {Wiescher}}]{schatz.aprahamian.ea:endpoint}
{Schatz}, H., {Aprahamian}, A., {Barnard}, V., {Bildsten}, L., {Cumming}, A.,
  {Ouellette}, M., {Rauscher}, T., {Thielemann}, F.-K., \& {Wiescher}, M. 2001,
  \prl, 86, 3471

\bibitem[{{Schatz} {et~al.}(2003a){Schatz}, {Bildsten}, \&
  {Cumming}}]{schatz.bildsten.ea:photodisintegration-triggered}
{Schatz}, H., {Bildsten}, L., \& {Cumming}, A. 2003a, \apjl, 583, L87

\bibitem[{Schatz {et~al.}(2003b)Schatz, Bildsten, Cumming, \&
  Ouellette}]{Schatz2003Nuclear-physics}
Schatz, H., Bildsten, L., Cumming, A., \& Ouellette, M. 2003b, \nphysa, 718, 247

\bibitem[{{Schatz} {et~al.}(1999){Schatz}, {Bildsten}, {Cumming}, \&
  {Wiescher}}]{schatz99}
{Schatz}, H., {Bildsten}, L., {Cumming}, A., \& {Wiescher}, M. 1999, \apj, 524,
  1014

\bibitem[{{Sedrakian} {et~al.}(2006){Sedrakian}, {M\"uther}, \&
  Schuck}]{SedrakianVertex-renormal}
{Sedrakian}, A., {M\"uther}, H., \& Schuck, P. 2006, ArXiv Astrophysics
  e-prints

\bibitem[{{Strohmayer} \& Bildsten(2004)}]{strohmayer03:review}
{Strohmayer}, T.~E. \& Bildsten, L. 2004, in Compact Stellar X-ray Sources, ed.
  W.~{Lewin} \& M.~{van der Klis} (Cambridge: Cambridge University Press), 113

\bibitem[{Strohmayer \& Brown(2002)}]{strohmayer.brown:remarkable}
Strohmayer, T.~E. \& Brown, E.~F. 2002, \apj, 566, 1045

\bibitem[{{Timmes} \& {Swesty}(2000)}]{timmes.swesty:accuracy}
{Timmes}, F.~X. \& {Swesty}, F.~D. 2000, \apjs, 126, 501

\bibitem[{{Tolman}(1939)}]{tolman:1939}
{Tolman}, R.~C. 1939, Physical Review, 55, 364

\bibitem[{{Wijnands}(2001)}]{wijnands:recurrent}
{Wijnands}, R. 2001, \apjl, 554, L59

\bibitem[{{Yakovlev} {et~al.}(2006){Yakovlev}, {Gasques}, {Afanasjev}, {Beard},
  \& {Wiescher}}]{Yakovlev2006Fusion-reaction}
{Yakovlev}, D.~G., {Gasques}, L.~R., {Afanasjev}, A.~V., {Beard}, M., \&
  {Wiescher}, M. 2006, \prc, 74, 035803

\bibitem[{{Yakovlev} {et~al.}(1999){Yakovlev}, {Kaminker}, \&
  {Levenfish}}]{yakovlev99:_neutr_cooper}
{Yakovlev}, D.~G., {Kaminker}, A.~D., \& {Levenfish}, K.~P. 1999, \aap, 343,
  650

\end{thebibliography}

\end{document}